\documentclass[sn-aps,pdflatex,10pt,twocolumn]{sn-jnl}% Style for submissions to Nature Portfolio journals
\usepackage{graphicx}%
\usepackage{multirow}%
\usepackage{amsmath,amssymb,amsfonts}%
\usepackage{amsthm}%
\usepackage{mathrsfs}%
\usepackage[title]{appendix}%
\usepackage{xcolor}%
\usepackage{textcomp}%
\usepackage{manyfoot}%
\usepackage{booktabs}%
\usepackage{algorithm}%
\usepackage{algorithmicx}%
\usepackage{algpseudocode}%
\usepackage{listings}%
\usepackage{url}
\usepackage{bm}
\renewcommand{\eqref}[1]{Eq.{$\left(\ref{#1}\right)$}}
\newcommand{\vecbr}[1]{\vec{\boldsymbol{\mathrm{#1}}}}

\theoremstyle{thmstyleone}%
\theoremstyle{thmstyletwo}%

\theoremstyle{thmstylethree}%

\begin{document}

\title[Article Title]{Levitated Ferromagnetic Torsional Oscillators for High-Precision Magnetometry and Probing Exotic Interactions}

 
\author[1]{\fnm{Yichong} \sur{Ren}}

\author[1]{\fnm{Lielie} \sur{Wu}}
 
\author[1]{\fnm{Wijnand} \sur{Broer}}

\author*[1]{\fnm{Fei } \sur{Xue}}\email{xfei.xue@hfut.edu.cn}

\author*[2]{\fnm{Pu} \sur{Huang}}\email{hp@nju.edu.cn}

\author[3,4,5,6]{\fnm{Jiangfeng} \sur{Du}}

\affil*[1]{\orgdiv{School of Physics}, \orgname{Hefei University of Technology}, \orgaddress{\city{Hefei}, \postcode{230601}, \state{Anhui}, \country{China}}}

\affil[2]{\orgdiv{National Laboratory of Solid State Microstructures and Department of Physics}, \orgname{Nanjing University}, \orgaddress{\city{Nanjing}, \postcode{210093}, \state{Jiangsu}, \country{China}}}

\affil[3]{\orgdiv{Institute of Quantum Sensing and School of Physics}, \orgname{Zhejiang University}, \orgaddress{\city{Hangzhou}, \postcode{310027}, \state{Zhejiang}, \country{China}}}
		
\affil[4]{\orgdiv{CAS Key Laboratory of Microscale Magnetic Resonance and School of Physical Sciences}, \orgname{University of Science and Technology of China}, \orgaddress{\city{Hefei}, \postcode{230026}, \state{Anhui}, \country{China}}}
		
\affil[5]{\orgdiv{Anhui Province Key Laboratory of Scientific Instrument Development and Application}, \orgname{University of Science and Technology of China}, \orgaddress{\city{Hefei}, \postcode{230026}, \state{Anhui}, \country{China}}}		

\affil[6]{\orgdiv{Hefei National Laboratory}, \orgname{University of Science and Technology of China}, \orgaddress{\city{Hefei}, \postcode{230088}, \state{Anhui}, \country{China}}}		


\abstract{
	Levitated ferromagnetic systems are expected to have significant potential in precision magnetic field sensing by leveraging mechanical isolation to minimize mechanical contact and associated noise. Here, we report the implementation of a high-sensitivity magnetometer based on a levitated ferromagnetic torsion oscillator, incorporating a centroid tracking method for superior measurement resolution and noises reduction. The device, featuring a compact sensor volume of $(2.5 \, \rm{mm})^3$ and operating under room temperature, attains a remarkable magnetic sensitivity of {$391\pm 59 \, \rm{fT\cdot Hz^{-1/2}}$}. This capability enables precise detection of weak magnetic fields and provides a novel platform for exploring exotic interactions beyond the Standard Model. These results demonstrate that the levitated torsion oscillator system not only serves as a powerful tool for high-precision magnetic sensing but also holds promise for advancing breakthroughs in fundamental physics.
}




\maketitle

\section*{Introduction}\label{sec1}

The precise measurement of magnetic fields is crucial in various scientific and technological domains, including geomagnetic navigation, biomagnetism, and the study of magnetic materials. In fundamental physics, high-sensitivity magnetic sensing plays a pivotal role in testing Lorentz symmetry \cite{pruttivarasin2015,brown2010,smiciklas2011}, searching for exotic spin-dependent interactions \cite{vasilakis2009,bulatowicz2013,lee2018,larry2013,rong2018,haowen2021,liang2023,wu2022}, detecting dark matter \cite{itay2022,abel2017,jiang2021,garcon2019,palken2020}, and measuring gravitational dipole moments \cite{fadeev2021g}. The sensitivity of magnetic sensors, however, is fundamentally limited by noises, primarily thermal and quantum detection noise in mechanical systems.
Specifically, thermal noise reduces sensitivity in accordance with the fluctuation-dissipation theorem, while quantum noise, governed by the Heisenberg uncertainty principle, establishes the standard quantum limit (SQL). 
In addition, the sensitivity of magnetic sensors is often limited by their spatial dimension, and ref.\cite{mitchell2020} provides a comprehensive overview of the relationship between magnetic noise and sensing volume, and furthermore summarizes the Energy Resolution Limits (ERL) based on available experimental results.

The detection of exotic spin-dependent interactions relies on measuring pseudomagnetic fields\cite{wu2022,liang2023}, with measurement accuracy predominantly governed by the sensor's magnetic sensitivity and the spatial separation between the magnetometer and the spin source. These parameters are typically subject to a trade-off relationship, where improved magnetic sensitivity generally demands a larger sensor volume. However, this increased volume inherently expands the sensor-source distance, thereby degrading measurement precision. Consequently, the ERL also serves as a critical metric for evaluating magnetometer performance in investigations of exotic interactions or potential fifth forces.

Ferromagnetic materials, characterized by a high density of correlated electron spins, are recognized for their strong potential in magnetic sensing applications \cite{vinante2021,kimball2016,barry2023}. In particular, ferromagnetic-mechanical systems, which detect magnetic forces or torques by measuring mechanical responses \cite{aireddy2019,colombano2020,gonzalez2021}, can leverage the full capabilities of ferromagnetic materials. These systems are typically constructed by attaching ferromagnetic materials to mechanical components like cantilevers, a method widely used in magnetic force microscopy and magnetic resonance force microscopy \cite{rugar2004,berman2006}. Ferromagnetic-mechanical experimental systems based on diamagnetic levitation\cite{ji2025} or on superconducting magnetic levitation \cite{ahrens2025} were realized, each demonstrating ultra-high magnetic sensitivity. This work introduces an additional experimental platform for ferromagnetic-mechanical systems, underscoring their promise in precision weak magnetic field detection and the search for novel physical interactions\cite{vinante2021}.

In this work, we experimentally demonstrate the capabilities of ferromagnetic torsion oscillators (FMTOs) with millimeter-scale dimensions operating at room temperature. Our results confirm the substantial potential of ferromagnetic materials for magnetic sensing, expanding the possibilities for fundamental physics research that relies on high-sensitivity magnetometers. Furthermore, we propose an FMTO-based magnetometer scheme to test exotic interactions and discuss its performance in detail.

\section*{Results}\label{sec2}
\subsection*{Magnetic sensing model of FMTO}
For a ferromagnet levitated in external magnetic fields $B_{\rm{bias}}$, the dynamic evolution is governed by
\begin{align}
	\dot{\vecbr{J}} &  =\vecbr{\tau}, \nonumber\\
	\dot{\vecbr{S}} &  =\vecbr{\Omega}\times \vecbr{S}. \label{eq00}
\end{align}
$\vecbr{J}=\vecbr{L}+\vecbr{S}$ is the total angular momentum, $\vecbr{L}=I\vecbr{\Omega}$ is the
kinetic angular momentum and $\vecbr{S}$ is the intrinsic spin, with the angular speed $\vecbr{\Omega}$ and moment of inertia $I$. $\vecbr{\tau}=\vecbr{\mu}\times \vecbr{B}$ is the torque on the ferromagnet from the magnetic field, $\vecbr{\mu}=\gamma_{\rm{0}}\vecbr{S}$ is the magnetic moment with the gyromagnetic ratio $\gamma_{\rm{0}}$.

The dominant dynamical behavior of the levitating ferromagnet depends on whether its intrinsic spin $(\vecbr{S})$ exceeds its momentum angular momentum $(\vecbr{L})$:  torsional dynamics prevail when spin dominance is pronounced \cite{vinante2021}, while gyroscopic effects dominate under angular momentum-dominant conditions \cite{fadeev2021f}. The magnitude of $\vecbr{L}$ and $\vecbr{S}$ are determined by ferromagnet's size and the bias magnetic field. For macroscopic ferromagnets and weak magnetic fields, the torsional pattern is dominant. In this case, the ferromagnet will librate like a torsional oscillator and is classified as an FMTO. Otherwise, the ferromagnet will precess around the external field like a gyroscope and is designated as a Ferromagnetic Gyroscopic Oscillator (FGO). An attempt to observe gyroscopic effects in a levitated ferromagnet was reported in Ref. \cite{vinante2025}

In our experiment, a cylindrical ferromagnet with an equivalent volume of $\bf{(2.5 \, \rm{mm})^3}$ is chosen to act as an FMTO, and the torsional motion is governed by
\begin{equation}
	I\ddot{\theta}+\gamma\dot{\theta}+k\theta=\tau_{\rm{s}}. \label{eq01}
\end{equation}
$\theta$ is the deflection angle in Fig.\ref{figs1}(b), $I$ is the moment of inertia of the FMTO, $\gamma=2\pi f_{\rm{r}}/Q$ is the dissipation, where $Q$ represents the quality factor and $f_{\rm{r}}=\sqrt{k/I}/2\pi$ is the resonant frequency. The torque elasticity coefficient $k\equiv \mu B_{\rm{bias}} $ can be tuned by the bias magnetic field $B_{\rm{bias}}$, which is aligned parallel to both the magnetic moment of the sensing magnet and the z-axis. This coefficient quantifies the restoring torque proportional to angular displacement. The driving torque $\tau_{\rm{s}}\left(  t\right)  =\mu  B_{\rm{sig}}\left(  t\right)$ acts perpendicular to both the sensing magnet’s magnetic moment and the z-axis, generating oscillatory motion in response to the magnetic signal $B_{\rm{sig}}$. $B_{\rm{sig}} \ll B_{\rm{bias}}$,  the orthogonal orientation ensures efficient torque transduction while minimizing parasitic coupling, as depicted in Fig. 1(a). The solution of \eqref{eq01} expressed in frequency domain is 
\begin{equation}
	\theta\left(  f\right)  =\chi\left(f\right)\tau_{\rm{s}}\left(  f\right).\label{eq06}
\end{equation}
$\tau_{\rm{s}}(f)$ is the spectrum of the driving torque $\tau_{\rm{s}}(t)$, $\chi\left(  f\right)=1/\left[4\pi^{2}I\left(f_{\rm{r}}^{2}-f^{2}+ i f f_{\rm{r}}/Q\right)\right]$ is the mechanical susceptibility of damped oscillator.

\subsection*{ Experimental details}
\begin{figure}[!tp]
	\centering
	\includegraphics[width=\linewidth]{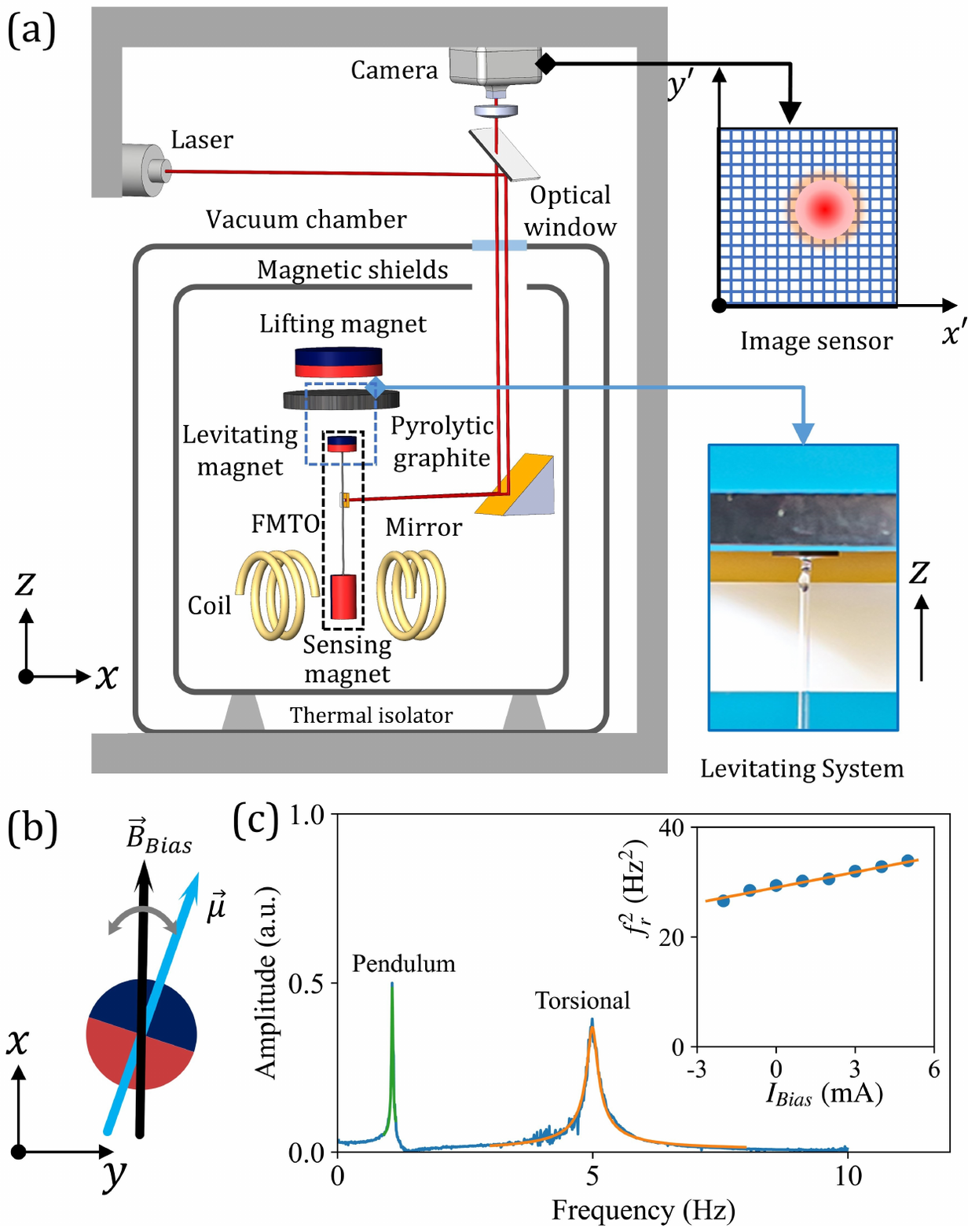}
	\caption{\label{figs1} 
			Diagram of the FMTO system. (a) FMTO consists a levitating-magnet, a glass rod and a sensing magnet. A pyroly graphite disk is used to stabilize the FMTO. The motion information of the FMTO can be extracted by the camera and centroid algorithm. The insert shows a typical gap of 0.5 mm between the lifting-magnet and the graphite disk. (b) The principle of the FMTO involves the introduction of a bias magnetic field and a signal magnetic field using two independent pairs of coils. (c) The linear fit depicted in the insert, illustrating the linear fit between the squared resonance frequency and the current in the DC coil, which confirms that  the orange resonant peak on the right side corresponds to the torsional mode of FMTO.}
	\end{figure}

The experimental setup is depicted in Fig.\ref{figs1}(a). The FMTO is housed within magnetic shields and a vacuum chamber at room temperature. 
The diamagnetic levitation apparatus is composed of a lifting magnet, a pyrolytic graphite, and a levitating magnet. FMTO maintain stable levitation through the interplay of its own gravity, the magnetic force exerted by the lifting magnet and the diamagnetic force exerted by the pyrolytic graphite on the levitation magnet, respectively.

This sensing magnet is cylindrical, made of NdFeB material, with a $1\,\rm{mm}$ diameter, $20\,\rm{mm}$ height, and magnetization parallel to its radial direction.
The bias field $B_{\rm{bias}}$ is introduced by DC current in the coils, while the magnetic signal $B_{\rm{sig}}$ is introduced by AC current in another AC coils. The mirror on the glass rod deflects the laser spot, thereby mapping the torsional motion of the FMTO to the translation of the laser spot on the image sensor. Subsequently, the motion information of the FMTO can be extracted by the camera and a centroid algorithm. The supporting structure is made of Teflon and is placed inside multilayer magnetic shields within a non-magnetic stainless steel vacuum chamber. Windows in the shields and vacuum chamber allow the laser beam to enter and exit. Thermal and vibrational isolations are also implemented throughout the system, including for the FMTO, magnetic shields, and measurement optics. An optimized centroid algorithm is employed to detect the torsional motion of the FMTO, effectively rejecting most common noise sources in the system. 
	Our measurement system, which integrates an optimized centroid tracking algorithm and post-processing digital lock-in technology, provides high-resolution, low-noise angle measurements with a wide dynamic range. This system is developed to explore the mechanical dynamic response characteristics of FMTO. The resulting responses are plotted in Fig.\ref{figs1}(c), where the resonance peaks at {$1.07\,\rm{Hz}$} and {$4.99\,\rm{Hz}$} correspond to the pendulum mode and the torsional mode of FMTO respectively. The torsional mode of FMTO, possessing a Q-factor of $Q=39.0\pm 0.4$ as determined by Lorentz fitting, will serve as the basis for magnetic sensing applications.
	According to the model of FMTO, the torsional elastic coefficient is determined by the bias magnetic field, which is indeed controlled by the bias current. Thus, the square of the resonant frequency should be proportional to the bias current $I_{\rm{bias}}$, allowing us to distinguish the torsional mode from the pendulum mode.  
	
	The inset of Fig.\ref{figs1}(c) illustrates the variation of the resonant frequency with the bias current, which confirms that the resonant peak at {$4.99\, \rm{Hz}$} corresponds to the torsional mode. Furthermore, a noticeable offset emerges in the fitting results depicted in the inset, suggesting that the elastic coefficient of FMTO exhibits a non-zero offset even when the bias magnetic field is set to zero.This non-zero offset in the resonant frequency likely stems from two sources: (1) rotational asymmetry in the magnetic field and structure of the FMTO, and (2) an asymmetrically structured FMTO elevating its center of mass during rotation. The latter effect creates a self-acting restoring torque, analogous to a tumbler’s dynamics.
	
	In experiments, we sequentially manufactured three FMTOs and utilized Lock-In method to measure their Q-factors, which ranged from $10$ to $100$, across different pressures. Subsequently, the FMTO with best overall performance was selected for magnetic sensing, showing a Q-factor of approximately $40$ at pressures below $10\, \rm{mbar}$ and less than $10$ at atmosphere.
	The primary sources of mechanical dissipation in FMTO systems are gas damping and magnetic dissipation. Gas damping, which is pressure-dependent, becomes negligible below $10\, \rm{mbar}$ in our setup, where magnetic  dissipation dominate. The latter arise primarily from eddy currents induced by time-varying magnetic fluxes in conductive materials, which is dominant under high vacuum conditions \cite{xie2023,ji2025}.
	
	Utilizing the composite materials, formed by blending small-particle diamagnetic materials with insulating substrates, in place of traditional bulk diamagnetic materials, can effectively reduce magnetic dissipation and enhance the Q-factor of diamagnetic levitation systems. By adopting this method and rationally optimizing the particle size, the Q-factor of diamagnetic levitation systems can be elevated to surpass $10^5$\cite{xiong2025,xie2023,ji2025}. In addition, recent technological advancements in torque sensing systems, such as superconducting levitation systems\cite{huillery2020,vinante2020,gieseler2020} and optical levitation systems \cite{ahn2020}, may offer alternative solutions for reducing dissipation and enhancing the Q-factor.

	\subsection*{Magnetic sensitivity calibration}

	Magnetic sensitivity, defined as the product of the minimum measurable magnetic field and the square root of the measurement time, is the most critical metric for evaluating magnetometer's performance. When adopting $\rm{SNR} \geqslant 1$ as the criterion for distinguishing signal from noises, the sensitivity is equivalent to the square root of the magnetic noise's power spectral density (PSD). 
	
	In our experiments, the raw data is the videos of the laser spot that captured by the camera, then the displacement signal is extracted from each frame of the video using the optimized centroid algorithm. The magnetic signal and the spot displacement signal are correlated through a transfer function, which is proportional to $\chi(f)$ in \eqref{eq06} (see Eq.S1 and Eq.S2 in Supplementary), and its square conforms to a Lorentz profile. After calibrating the transfer function, magnetic sensitivity is derived by measuring the noise PSD. The calibration process, detailed in Supplementary Method 1, involves applying a magnetic signal with predefined amplitude and frequency to drive the FMTO, then measuring the resulting optical spot displacement response signal. This generates a calibration peak in the response signal’s PSD, and the transfer function at the driving frequency is determined using the ratio of the peak’s root mean square amplitude to the applied magnetic signal amplitude. By integrating this result with resonance frequency and Q-factor data obtained via Lock-In detection, the transfer function is calibrated across the full frequency range. Finally, magnetic sensitivity is calculated by combining the noise PSD with the calibrated transfer function, following Eq. S3 in the Supplementary Materials.
	
	\begin{figure}[tp]
	\centering
	\includegraphics[width=\linewidth]{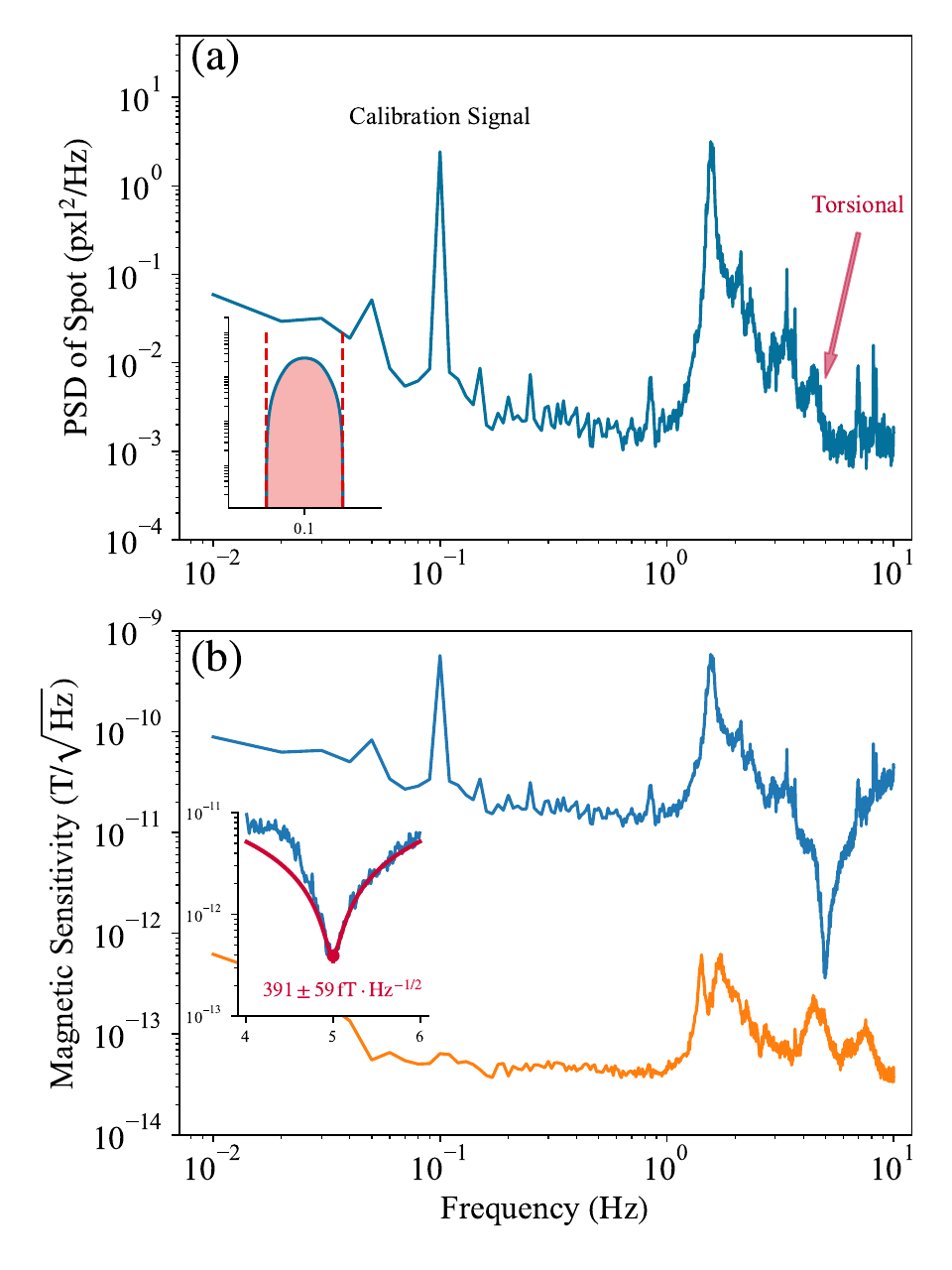}% Here is how to import EPS art
	\caption{\label{figs3} Calibration of magnetic sensitivity. (a) The PSD of the laser spot's movement on the CCD is measured while the FMTO is driven by a magnetic signal. (b) The magnetic sensitivity is derived using the calibration magnetic signal. The blue line represents the PSD of the FMTO without any driving. The inset illustrates the relationship between magnetic sensitivity and frequency, particularly near torsional resonance. While the orange line represents the noise floor of the FMTO, which is limited by the measurement noise.}
	\end{figure}
	
	As illustrated in Fig.\ref{figs3}(a), a  magnetic calibration signal at $0.1\, \rm{Hz}$ is generated through the AC coils by an sinusoidal current, the generation and calibration of this magnetic signal is detailed in Supplementary Method 2. The transfer function is determined by combining these results with the mechanical characteristics of the FMTO, including its resonant frequency and $Q$-factor. To demonstrate the performance of our FMTO system in probing magnetic fields, we consider two scenarios. In the first scenario, we determine the minimum measurable magnetic field based on the FMTO's PSD, represented by the blue line in Fig.\ref{figs3}(b). Under resonant conditions, a sensitivity of {$391 \pm 59\,\rm{fT\cdot Hz^{-1/2}}$}  is achieved at {$4.99\,\rm{Hz}$}.
	Fig.\ref{figs3}(a) reveals a minor peak at the resonant frequency in the noise PSD, while Fig.\ref{figs3}(b) exhibits a corresponding dip in magnetic sensitivity at the same frequency. This behavior arises from resonant amplification in the FMTO: at resonance, the system amplifies mechanical responses, enhancing sensitivity to weak magnetic signals. However, resonance also amplifies mechanical noise, manifesting as a PSD peak. While this amplification improves detection of weak signals, the simultaneous noise enhancement obscures finer magnetic details, resulting in a notable dip in sensitivity at the resonant frequency.

	In the second scenario, the FMTO is fixed onto the graphite disk to inhibit its torsional motion. This allows the obtained PSD to represent the optical measurement's noise floor limit (the detailed analysis of system noise refers to Supplementary Analysis 1), yielding a measurement limited sensitivity of {$55 \, \rm{fT\cdot Hz^{-1/2}}$} around $0.1 \,\rm{Hz}$ , the orange line in Fig.\ref{figs3}(b).

	\begin{figure}[tp]
		\centering
		\includegraphics[width=1.0\linewidth]{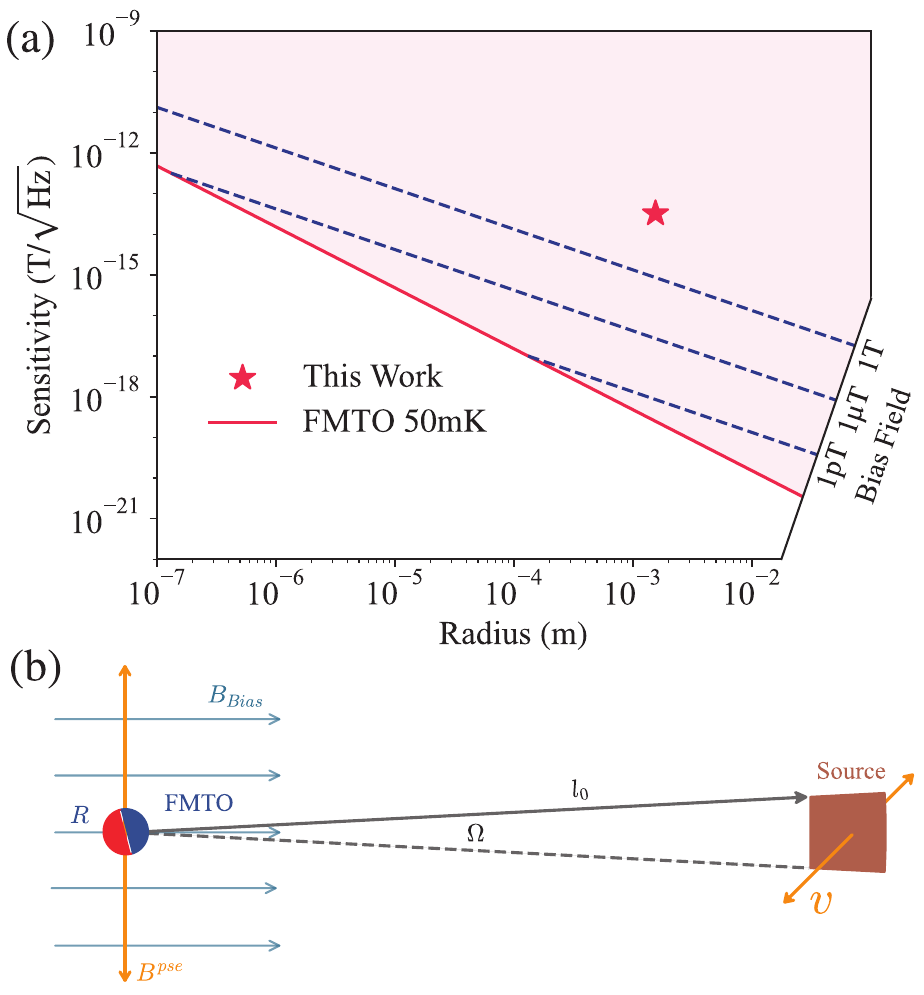}% Here is how to import EPS art
		\caption{\label{figs6} Optimal Magnetic sensitivity under typical bias field for versus the sensor's radius. $(a)$ The potential sensitivity of FMTO magnetometer versus the sensor's radius. $(b)$ FMTO magnetometer for probing the pesudo-magnetic field generated by vibrating nucleon source.} 
	\end{figure}
	
	\section*{Discussion}
	\subsection*{Magnetic sensing potential of FMTO}
	For an FMTO magnetometer, the sensitivity limited by thermal noise is given as:
	\begin{equation}
\eta_{B_{\rm{th}}}\left(  f\right)  = S_{\tau_{\rm{th}}}^{1/2}\left(  f\right)
/\mu, \label{eq10}
	\end{equation} where $S_{\tau_{\rm{th}}}\left(  f\right)  =4k_{\rm{B}}TI\gamma$ is the PSD of thermal torque noise, and $\mu$ is the magnetic moment. According to \eqref{eq10}, the potential capacity of FMTO with a Q-factor of $10^{5}$ at $50\,\rm{mK}$ is illustrated with red region in Fig.\ref{figs6}(a), see the detailed derivation in Supplementary Analysis 2. The sensitivity achieved under different bias fields is shown by the blue dashed line, which indicates that for FMTO systems, a weaker bias field or a larger sensor radius can improve the sensitivity.
	The sensitivity of \textcolor{red}{$55\,\rm{fT\cdot Hz^{-1/2}}$} is attainable by our FMTO system with an equivalent radius of $1.5\,\rm{mm}$ at room temperature, as indicated by the red star in the Fig.\ref{figs6}(a). 
	One of the promising applications of the FMTO system is probing pseudo-magnetic fields arising from exotic interactions beyond the standard model \cite{fadeev2021f,vinante2021}. Fig.\ref{figs6}(b) introduces a proposal to test exotic interactions using an FMTO magnetometer. A vertical pseudo-magnetic field is generated by a nucleon source vibrating perpendicular to the page. The FMTO magnetometer, positioned at a distance $l_{\rm{0}}$ from the source, is used to detect this pseudo-magnetic field. The nucleon source consists of a spherical shell-shaped BGO crystal with a cubic angle of $\Omega$.

	\subsection*{ Probing exotic interactions}
	For a hypothetical new-bosons-mediated spin- and velocity-dependent interaction, the vertical pseudomagnetic field generated by a vibrating nucleon source can be expressed by\cite{wu2023}
	\begin{equation}
		\vecbr{B}^{\rm{pse}}=f^{4+5}C_{\rm{0}}\int%
		_{V}dV\left[  \left(  \vecbr{v}_{\rm{n}}\times\hat{r}\right)  \left(  \frac{1}{\lambda
			r}+\frac{1}{r^{2}}\right)  e^{-r/\lambda}\right] ,  \label{eq09}%
	\end{equation}
	where $\vecbr{v}$ is the relative velocity,$\hat{r}=\vecbr{r}/r$ is	the unit vector,$r$ is the distance between the electron spin and nucleon, $f^{4+5}=g_{\rm{s}}^{\rm{e}}g_{\rm{s}}^{\rm{N}}$ is the dimensionless coupling constant, $\lambda$ is the Compton wavelength of the unobserved boson, and $C_{\rm{0}}$ is a parameter associated with materials and fundamental constants.
	
	Assuming the source amplitude $A_{\rm{n}}$ and FMTO radius $R$ are much smaller than the distance $l_{\rm{0}}$, we define $\varepsilon_{\rm{A}}=A_{\rm{n}}/l_{\rm{0}}\ll 1$. If the source vibrates at frequency $f_{\rm{n}}$, the amplitude of its velocity is $v_{\rm{n}}=2\pi f_{\rm{n}} A_{\rm{n}}=2\pi f_{\rm{n}} \varepsilon_{\rm{A}} l_{\rm{0}}$. The pseudo-magnetic field generated by the total source in Fig.\ref{figs6}(b) can be calculated by integrating \eqref{eq09} after substituting the velocity: 	
	\begin{equation}
		\vecbr{B}^{\rm{pse}}=2\pi C_{\rm{0}}C_{\lambda}\left(  l_{\rm{0}}\right)f^{4+5}f_{\rm{n}}\Omega
		\varepsilon_{\rm{A}}\lambda^{2},  \label{eq09a}
	\end{equation}
	where $C_{\lambda}(l_{\rm{0}})=(l_{\rm{0}}^2/\lambda^{2}+2l_{\rm{0}}/\lambda)e^{-l_{\rm{0}}/\lambda}$ is a dimensionless function with respect to $\lambda$ and $l_{\rm{0}}$.
	
	If an FMTO magnetometer with sensitivity $\eta_{\rm{B}}(f)$ is used to test exotic interactions by probing the pseudo-magnetic field over a time $T_{\rm{mea}}$, the resulting dimensionless coupling constant is given by
	\begin{equation}
		\Delta f^{4+5}=\frac{\Delta B}{\left\vert \partial B^{\rm{pse}}/\partial
			f^{4+5}\right\vert }=\frac{\eta_{\rm{B}}\left(  f\right)  T_{\rm{mea}}^{-1/2}}%
		{2\pi\varepsilon_{\rm{A}}\Omega C_{\rm{0}}C_{\lambda}\left(  l_{\rm{0}}\right)  f_{\rm{n}}%
			\lambda^{2}},\label{eq11}%
	\end{equation}%
	$\Delta B=\eta_{\rm{B}}(f) T_{\rm{mea}}^{-1/2}$ is the minimum detectable magnetic field of FMTO magnetometer. 
	
	According to \eqref{eq09a}, the pseudo-magnetic signal decays rapidly with increasing source-sensor distance $l_{\rm{0}}$ beyond a critical threshold, limiting the sensor's dimensions and sensitivity. The precision of probing exotic interactions relies on the sensor's magnetic sensitivity within a confined space. As noted in Ref.~\cite{vinante2021}, FMTO, a ferromagnetic sensor, exhibits significantly higher sensitivity than other magnetic sensors of equivalent volume, making it well suited for probing exotic interactions.

	\begin{figure}[bp]
		\centering
		\includegraphics[width=1.0\linewidth]{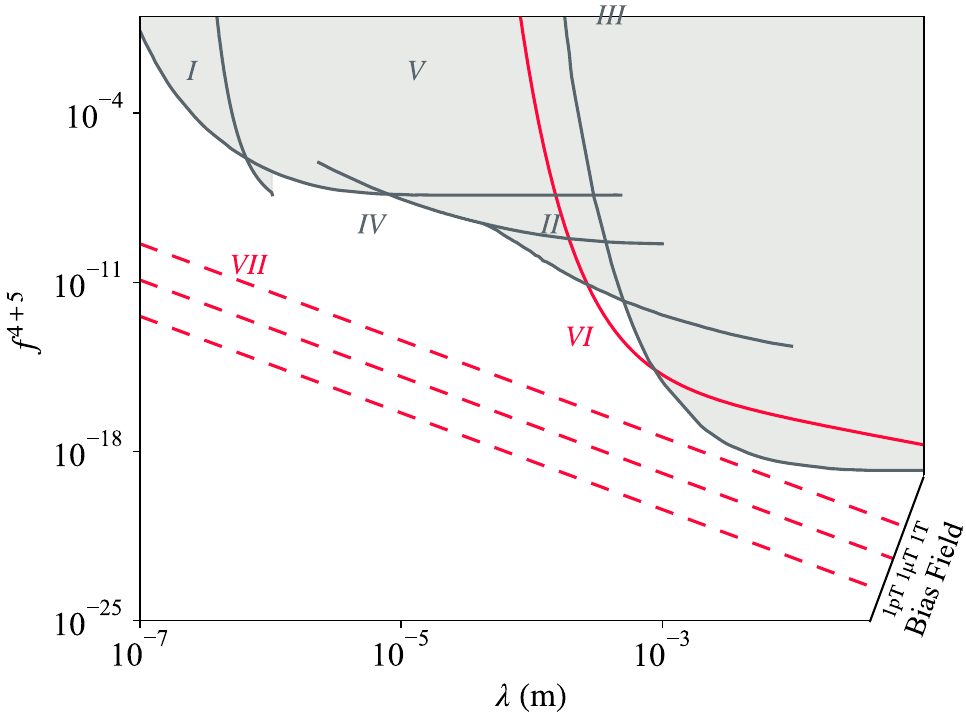}% Here is how to import EPS art
		\caption{\label{figs4} The best bound of coupling constant $f^{4+5}$.  experimental result are shown as shaded in gray color above solid lines: I.\cite{ding2020}, II.\cite{wu2023}, III.\cite{diguang2023}, IV.\cite{piegsa2012}, V.\cite{kim2018}. The red solid line VI gives the best bound for the FMTO realized in this work with optimal setup. The red dashed lines plot the upper bounds set by optimized FMTO magnetometers, limited by thermal torque noise, for various bias magnetic fields. }
	\end{figure}
	
	If the FMTO magnetometer, with a sensitivity of \textcolor{red}{$55\,\rm{fT\cdot Hz^{-1/2}}$}
	, is utilized to probe the exotic interactions, the results are depicted by the red solid line in Fig.\ref{figs4}.
	Additionally, the other parameters are configured as follows: $f_{\rm{n}}=4.99\,\rm{Hz},l_{\rm{0}}=7.5\,\rm{mm},A_{\rm{n}}=1.5\,\rm{mm},\Omega=10^{-2}\,\rm{sr},T_{\rm{mea}}=10^{4} \, \rm{s}$. 
	Next, we evaluate the potential of all FMTO magnetometers in probing the exotic interactions mediated by the axions with a force range of $\lambda$. In this process, the distance $l_{\rm{0}}$ will be fixed at the optimal value $\sqrt{2}\lambda$ and the source amplitude will be set to $A_{\rm{n}}=\varepsilon_{\rm{A}}l_{\rm{0}}=\sqrt{2}\varepsilon_{\rm{A}}\lambda$. The other experimental parameters are set as $T=50\,\rm{mK},Q=10^{7},T_{\rm{mea}}=10^{4} \, \rm{s}$. The red dashed line in Fig.\ref{figs4} shows the results obtained under different bias field conditions. Increasing the bias field reduces the sensitivity of FMTO, but simultaneously increases the resonant frequency, resulting in a stronger pseudo-magnetic signal and ultimately improving the probing accuracy of exotic interactions (see Eq.S10 in Supplementary Analysis 3 for details).

	The FMTO magnetometer, with extremely high magnetic sensitivity, demonstrates remarkable capability in probing exotic interactions, in particular those that are mediated by massive axions characterized by their short Compton wavelengths. By optimizing parameters like the source-to-magnetometer distance, source amplitude, and Q-factor, it enhances sensitivity, improving upper bounds on the dimensionless coupling constant $f^{4+5}$ by many orders of magnitude over existing methods. Its adjustable properties, including resonant frequency and bias field, enable precise measurements across various conditions, surpassing current techniques and facilitating exploration of new physics and fundamental interactions.
	
    \section*{Conclusion}
	
	In conclusion, the FMTO system offers notable advantages for weak magnetic field measurements. Sensitivity can be further enhanced by increasing the Q-factor, radius, or decreasing the temperature. For example, a sensitivity of $0.23 \, \rm{aT\cdot Hz^{-1/2}}$ at $1 \, \rm{mHz}$  can be achieved with $Q=10^{5},T=50\,\mathrm{mK}, r=10\,\rm{mm}$ ($r$ is the radius of ferromagnet sphere). Unlike the system proposed in \cite{vinante2021}, which relies solely on the ferromagnetic oscillator’s size, our FMTO system allows resonant frequency adjustment via the bias magnetic field. This provides greater flexibility for targeting specific interaction frequencies, potentially revealing new physics. Moreover, the ability to perform measurements at room temperature enhances the practicality of the FMTO for a wide range of experimental conditions, providing a more accessible tool for researchers exploring fundamental interactions.

	\section*{Methods}
	\subsection*{Measurement system} 
	An optimized centroid algorithm is employed to extract motion information from the laser spot captured in the camera's video. Taking advantage of this algorithm, the camera offers an excellent solution for detecting the deflection of FMTO. This optimized centroid algorithm enables high-resolution measurement of the torsional angle with a standard deviation of less than $2 \times 10^{-8} \, \rm{rad}$. It also exhibits a linear response to the torsional motion from $10^{-8} \, \rm{rad}$ to $10^{-3} \, \rm{rad}$. It demonstrates high precision, a wide dynamic range, and the capability to reduce various noise sources, including electrical noise in the laser source and image sensor, as well as the laser spot drifting caused by deformation of the system's supporting frames. Such deformation may stem from temperature fluctuations or vibrations in the components along the optical path.

	Meanwhile, time fluctuations in each video frame recorded by the camera can lead to phase fluctuations during lock-in measurements, compromising accuracy. To tackle this challenge, a digital post-processing lock-in technique has been developed and implemented in our experiments. This technique effectively mitigates the random delays introduced by the image sensor and data processing.

	Together, these efforts—the optimal centroid algorithm for laser spot detection and the digital post-processing lock-in technique to address random delays from the camera—enable high-accuracy, low-noise measurement of the torsional motion of the FMTO.

\backmatter

\section*{Data availability}
Data for all figures in the main text and supplemental material is available from the corresponding authors upon reasonable request.

\bibliography{sn-bibliography}% common bib file

\section*{Acknowledgements}
We thank Prof. Xing Rong for his valuable input on the writing. This work was supported by the Fundamental Research Funds for the Central Universities Grant No. JZ2025HGTB0176, the National Key R\&D Program of China Grant No.2024YFF0727902, the National Natural Science Foundation of China Grant No. 12150011, No.T2388102, and No.12075115.

\section*{Author contributions}
All authors contributed to the development and writing of this paper. The general idea was conceived by F.X., P.H., and J.D.; the experimental setup was designed by F.X. and P.H., and subsequently built by Y.-C. R. and F.X. Experimental measurements were performed by L.-L. W. and Y.-C. R., with Y.-C. R. also conducting the theoretical calculations. F.X. supervised the work. The manuscript was written by F.X., Y.-C. R., and W. B., with all authors providing discussions and inputs throughout the process.

\section*{Competing interests}

The authors declare no competing interests.

\section*{Additional information}

\noindent{\bf Supplementary information} The online version contains supplementary material available.

\noindent{\bf Correspondence} and requests for materials should be addressed to F. X. or P. H.

\end{document}

\title[Article Title]{Supplementary Information for "Levitated Ferromagnetic Torsional Oscillators for High-Precision Magnetometry and Probing Exotic Interactions"}

\author[1]{\fnm{Yichong} \sur{Ren}}

\author[1]{\fnm{Lielie} \sur{Wu}}

\author[1]{\fnm{Wijnand} \sur{Broer}}

\author*[1]{\fnm{Fei } \sur{Xue}}\email{xfei.xue@hfut.edu.cn}

\author*[2]{\fnm{Pu} \sur{Huang}}\email{hp@nju.edu.cn}

\author[3,4,5,6]{\fnm{Jiangfeng} \sur{Du}}

\affil*[1]{\orgdiv{School of Physics}, \orgname{Hefei University of Technology}, \orgaddress{\city{Hefei}, \postcode{230601}, \state{Anhui}, \country{China}}}

\affil[2]{\orgdiv{National Laboratory of Solid State Microstructures and Department of Physics}, \orgname{Nanjing University}, \orgaddress{\city{Nanjing}, \postcode{210093}, \state{Jiangsu}, \country{China}}}

\affil[3]{\orgdiv{Institute of Quantum Sensing and School of Physics}, \orgname{Zhejiang University}, \orgaddress{\city{Hangzhou}, \postcode{310027}, \state{Zhejiang}, \country{China}}}

\affil[4]{\orgdiv{CAS Key Laboratory of Microscale Magnetic Resonance and School of Physical Sciences}, \orgname{University of Science and Technology of China}, \orgaddress{\city{Hefei}, \postcode{230026}, \state{Anhui}, \country{China}}}

\affil[5]{\orgdiv{Anhui Province Key Laboratory of Scientific Instrument Development and Application}, \orgname{University of Science and Technology of China}, \orgaddress{\city{Hefei}, \postcode{230026}, \state{Anhui}, \country{China}}}		

\affil[6]{\orgdiv{Hefei National Laboratory}, \orgname{University of Science and Technology of China}, \orgaddress{\city{Hefei}, \postcode{230088}, \state{Anhui}, \country{China}}}		


\abstract{
This supplementary document further elaborates on the measurement method, calibration of magnetic sensitivity, noise-related issues, and the magnetic sensitivity limited by thermal noise for levitated ferromagnetic torsional oscillator. Analytical expressions for the integral form of the pseudomagnetic field are provided, along with the results of probing exotic interactions. Additionally, material parameters for ferromagnetic and nucleon materials are presented, offering a comprehensive overview of the experimental setup and theoretical underpinnings.}

\maketitle

\section*{Supplementary Method 1: magnetic calibration}

Magnetic sensitivity calibration primarily relies on the accurate determination of the transfer function between magnetic signals and spot displacement. Given that the square of this transfer function adheres to a Lorentzian profile, the calibration process comprises two key steps: First, determine the resonance frequency and Q-factor of the FMTO to establish the foundational shape of the transfer function; second, perform precise calibration of the transfer function at a specific frequency. By merging these outcomes, a comprehensive calibration of the transfer function across the entire frequency spectrum can be achieved. Subsequently, the magnetic sensitivity of the FMTO can be computed using the transfer function and the noise power spectral density (PSD).

\begin{figure}[tbp]
	\centering
	\includegraphics[width=0.4\linewidth]{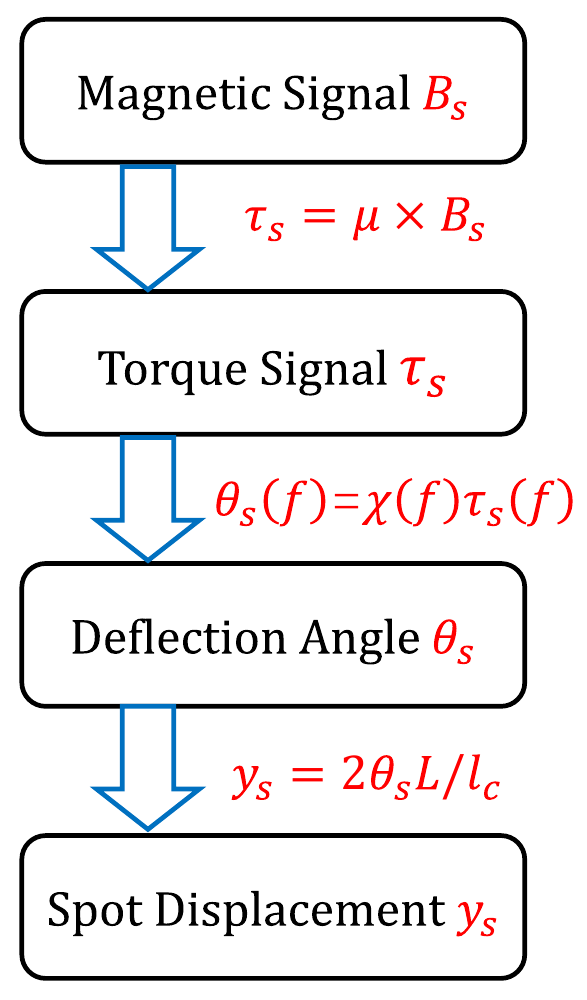} 
	%Here is how to import EPS art
	\caption{Logic and procedure for calibrating the coefficients in tranducing magnetic singals into pixels of laser spot displacement.}
	\label{figs1}
\end{figure}

Fig.\ref{figs1} illustrates the transfer process in experiments: from magnetic signals to pixels. The transformation function relating the magnetic signal to the laser spot's displacement can be represented as:
\begin{equation}
	y(f) = C_{\rm{T}}(f) B_{\rm{s}}(f). \label{s01}
\end{equation}
Here, $B_{\rm{s}}(f)$ denotes the spectrum of magnetic signal, $y(f)$ represents the spectrum of of spot displacement in pixels, and $C_{\rm{T}}(f)$ is the transfer function. The theorical expression of transfer function $C_{\rm{T}}(f)$ is:
\begin{equation}
	C_{\rm{T}}(f) = 2\mu L \chi(f)/l_{\rm{c}}, \label{s02}
\end{equation}
where $\mu$ is the magnetic moment of FMTO, $L$ is the path length of the laser from the mirror to the camera, $l_{\rm{c}}$ is the pixel size of the camera, and $\chi(f)$ is the mechanical sensitivity coefficient of the FMTO.

{To calibrate the transfer function $C_{\rm{T}}(f)$, we applied a sinusoidal current with a frequency of $0.1\,\rm{Hz}$ and an amplitude of $2\,\rm{\mu A}$ in the AC coils. This current generated a magnetic field with an amplitude of $114\pm 17 \,\rm{pT}$ to drive the FMTO. This known reference signal produces a prominent peak in the PSD of the FMTO response signal, as shown by the calibration signal in Fig.\ 2(a). The PSD in Fig.\ 2(a) was obtained by dividing the time-series data into multiple $100\,\rm{s}$ segments, selecting the $50$ segments with the highest signal-to-noise ratio (SNR), and averaging the PSDs for these segments.}

{By calculating area $A_{\rm{cal}}$ encompassed by this calibration peak  and the amplitude $B_{\rm{cal}}$ of the calibration signal , the transfer function $C_{\rm{T}}(f)$ at $0.1\,\rm{Hz}$ is determined:  $C_{\rm{T}}\left( 0.1\right) =\sqrt(2A_{\rm{cal}})/B_{\rm{cal}}=2.7\pm 0.4\,\rm{pxl/fT}$. }
Note that $|C_{\rm{T}}\left( f\right)|^{2}$ conforms to a Lorentzian profile because $|C_{\rm{T}}\left( f\right)|\propto |\chi\left(f\right)|$. Thus the outline of $|C_{\rm{T}}\left( f\right)|^{2}$ is determined by the Q-factor and the resonance frequency, both of which can be measured using the Lock-In method. A comprehensive scan of the frequency response and phase response of the FMTO is conducted over the range from $0.01\ \rm{Hz}$ to $10\ \rm{Hz}$, with a step size of $0.01\ \rm{Hz}$. Subsequently, a Lorentzian fit was then conducted on the squared amplitude, thereby confirming a resonant frequency of $4.99\,\rm{Hz}$ and a Q-factor of $39.0\pm0.4$. The transfer functions over entire frequency can then be derived from this result with aforementioned $C_{\rm{T}}\left( 0.1\right) $, thus the magnetic sensitivity (or magnetic noise floor) can be obtained by dividing the spot's noise floor by the transfer function $C_{\rm{T}}\left( f\right) $.

The magnetic sensitivity $\eta_{\rm{B}}(f)$ is therefore calculated by using the measured noise floor $S_{y}^{1/2}(f)$ of the laser spot displacement and the transfer function $C_{\rm{T}}(f)$:
\begin{equation}
	\eta_{\rm{B}}(f) = S_{y}^{1/2}(f)/C_{\rm{T}}(f).
	\label{s03}
\end{equation}

\section*{Supplementary Method 2: magnetic signal generation and detailed parameters}
{The current-magnetic conversion coefficient of the coils was theoretically calculated based on the Biot–Savart law. To verify the validity of the theoretical model, a pair of DC coils was experimentally characterized. The coils were wound with copper wire, each consisting of $80$-turns with a diameter of $60\,\rm{mm}$ and a separation of $38\,\rm{mm}$. A $9\,\rm{V}$ battery combined with a variable resistor was used to generate DC currents ranging from $1\,\rm{mA}$ to $60\,\rm{mA}$. The magnetic field at the coil center was measured using a commercial magnetometer (\href{http://www.sbjd88.com/products/tm4300.html}{TM4300B}, resolution 10 nT). The measured current-magnetic conversion coefficient($1.73\ \rm{\mu T/mA}$) was approximately $14.4\%$ lower than the theoretical value($2.02\ \rm{\mu T/mA}$), indicating a systematic deviation likely due to geometric tolerances and experimental uncertainties.}

{The AC coils were also wound from copper wire, forming a pair of $5$-turns coils with a diameter of $50\,\rm{mm}$ and a separation of $65\,\rm{mm}$. Based on the Biot–Savart law, the theoretical current–magnetic field conversion coefficient was calculated to be $56.95\ \rm{nT/mA}$. Considering the discrepancy observed in the DC coil calibration, this value was adopted with an assumed uncertainty of $±15\%$ for subsequent calibration.} In experiments, the AC coil are serially connected to a $1\ \rm{M\Omega}$ resistor and driven by the data acquisition card, specifically the \href{https://www.ni.com/docs/en-US/bundle/usb-6216-specs/page/specs.html}{NI USB-6216} model. This card has a maximum output voltage of $\pm 10\ \rm{V}$ and a maximum current of $2\ \rm{mA}$.

For ferromagnetic materials like $\rm{NdFeB}$ \cite{vinante2020,fadeev2021f}, the mass density $\rho_{\rm{m}} = 7430\,\rm{kg/m^{3}}$, the magnetization $\mu_{\rm{0}}M = 0.71\,\rm{T}$, and the electron spin density $\rho_{\rm{e}} = 6 \times 10^{28}\,\rm{m^{-3}}$. For $\rm{Bi_{4}Ge_{3}O_{12}\,(BGO)}$ materials \cite{kim2018}, the density of nucleons $\rho_{\rm{n}} = 4 \times 10^{30}\,\rm{m^{-3}}$. The glass rod is a hollow cylinder with an inner diameter of $0.8\,\rm{mm}$, an outer diameter of $1\,\rm{mm}$, and a length of $20\,\rm{cm}$. The path length of the laser is approximately $0.7\,\rm{m}$. Both sensing and levitating magnets are manufactured using $\rm{NdFeB}$ material. The sensing magnets are cylindrical in shape, with a diameter of $1\,\rm{mm}$ and a length of $20\,\rm{mm}$, while the levitating magnets are disc-shaped, with a diameter of $5\,\rm{mm}$ and a thickness of $0.5\,\rm{mm}$. The mirror glued to the center of the glass rod is made of a $3\,\rm{mm} \times 5\,\rm{mm}$ thin silicon wafer. The total mass of the torsional oscillator is approximately $400\,\rm{mg}$, with a moment of inertia of approximately $3 \times 10^{-10}\,\rm{kg \cdot m^2}$.

\section*{Supplementary Analysis 1: noise source}

For sensitive magnetic field measurements, it is essential to understand these noises from the environment and the measurement process. To this end, we analyzed the following four noise sources that are most relevant:
\begin{itemize}
	\item Magnetic noise in the experimental environment, which requires a proper strategy to reduce its influence;
	\item Measurement noise of the laser spot displacement detection, which sets the limit of the measurement method;
	\item Vibration noise transducing on the laser spot displacement, which add noise both on the dynamics of FMTO and on laser spot displacements.
	\item Thermal noise, which determines the limit of the magnetic sensitivity of the FMTO at finite temperature.
	
\end{itemize}

\begin{figure}[tp]
	\centering
	\includegraphics[width=\linewidth]{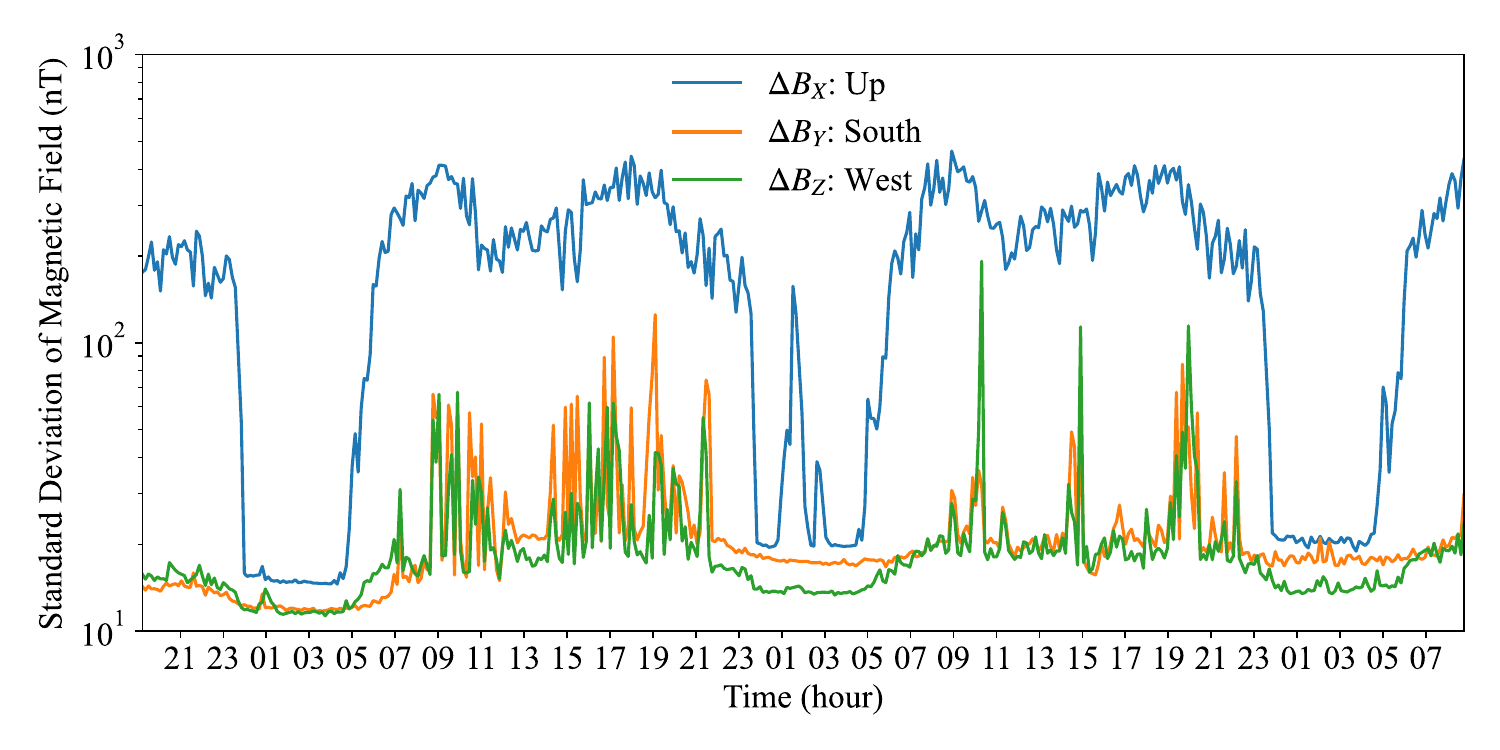} 
	%Here is how to import EPS art
	\caption{The environmental magnetic field fluctuations in the experiment.}
	\label{figs2}
\end{figure}

{\bf Magnetic noise.} In experiment, environmental magnetic noise can sesiously influence the magnetic sensitivity of FMTO. 
We have measured the magnetic fields in the laboratory using a magnetometer \href{https://www.pnisensor.com/rm3100/} {RM3100} and calculated their deviations. The typical numbers are plotted in Fig.\ref{figs2}, showcasing a standard deviation of approximately $10\ \text{nT}$ in the horizontal direction. In the vertical direction, the standard deviation varies significantly, ranging from $20\ \text{nT}$ during the period from 00:00 to 05:00 to $200\ \text{nT}$ from 06:00 to 24:00. This notable difference in magnetic noise along the vertical direction across different time periods is attributed chiefly to the influence of the metro. During the daytime (06:00 to 24:00), the metro generates a substantial amount of vertical magnetic noise, which decreases at night when the metro stops operating (00:00 to 05:00). Ref.~\cite{zhang2018} also reported measurements of the geomagnetic field in Beijing, revealing a significant impact of the metro on vertical magnetic noise.

Geomagnetic monitoring data gathered by geomagnetic stations serve as a basis for estimating the geomagnetic noise floor. Specifically, Ref.~\cite{guo2024} indicates that in the Huainan and Mengcheng regions, the geomagnetic noise floor are less than $10\ \rm{pT\cdot Hz^{-1/2}}$ at $0.01\ \rm{Hz}$. Similarly, Ref.~\cite{zhang2018} states that at the Hongshan station, the noise floor is around $30\ \rm{pT\cdot Hz^{-1/2}}$ at $0.5\ \rm{Hz}$, derived from the square root of the PSD. Other literature \cite{Emine2017,Krylov2014} does not provide the PSD of the geomagnetic field, yet the standard deviations they report are consistent with those presented in Refs.~\cite{zhang2018,guo2024}.

The magnetometer \href{https://www.pnisensor.com/rm3100/} {RM3100} only has an LSB (Least Significant Bit) of $13\ \rm{nT}$, indicating that the measured standard deviation of $10\ \rm{nT}$ may not be accurate. The actual standard deviation is likely to be substantially smaller than this measured value. Due to the lower precision of the \href{https://www.pnisensor.com/rm3100/} {RM3100}, it is not feasible to accurately determine the fluctuations and PSD (Power Spectral Density) of the magnetic noise in the laboratory at night. 

Therefore, we conclude that when the metro is not running at night, the geomagnetic noise floor can be as low as the order of $10\ \rm{pT\cdot Hz^{-1/2}}$. Furthermore, multi-layer magnetic shielding can reduce external magnetic noise by $3$ to $5$ orders of magnitude. Thus, based on the aforementioned knowledge of the ambient magnetic field, the experimental data were collected remotely during quiet nights.

{\bf Measurement noise.} A \href{https://www.hikrobotics.com/en/machinevision/productdetail?id=5763}{Hikvision MV-CA050-20UM} camera, featuring a pixel size of $4.8\ \rm{\mu m} \times 4.8\ \rm{\mu m}$, operating in 8-bit mode, is adopted to capture the spot's displacement at a frame rate of $50\ \rm{FPS}$. The power spectral density (PSD) of the quantization noise introduced by this camera is approximately $8 \times 10^{-7}\,\rm{pxl^{2}\cdot Hz^{-1}}$, 
which is significantly lower than the thermal noise of the FMTO under non-resonant conditions.

\begin{figure}[bp]
	\centering
	\includegraphics[width=\linewidth]{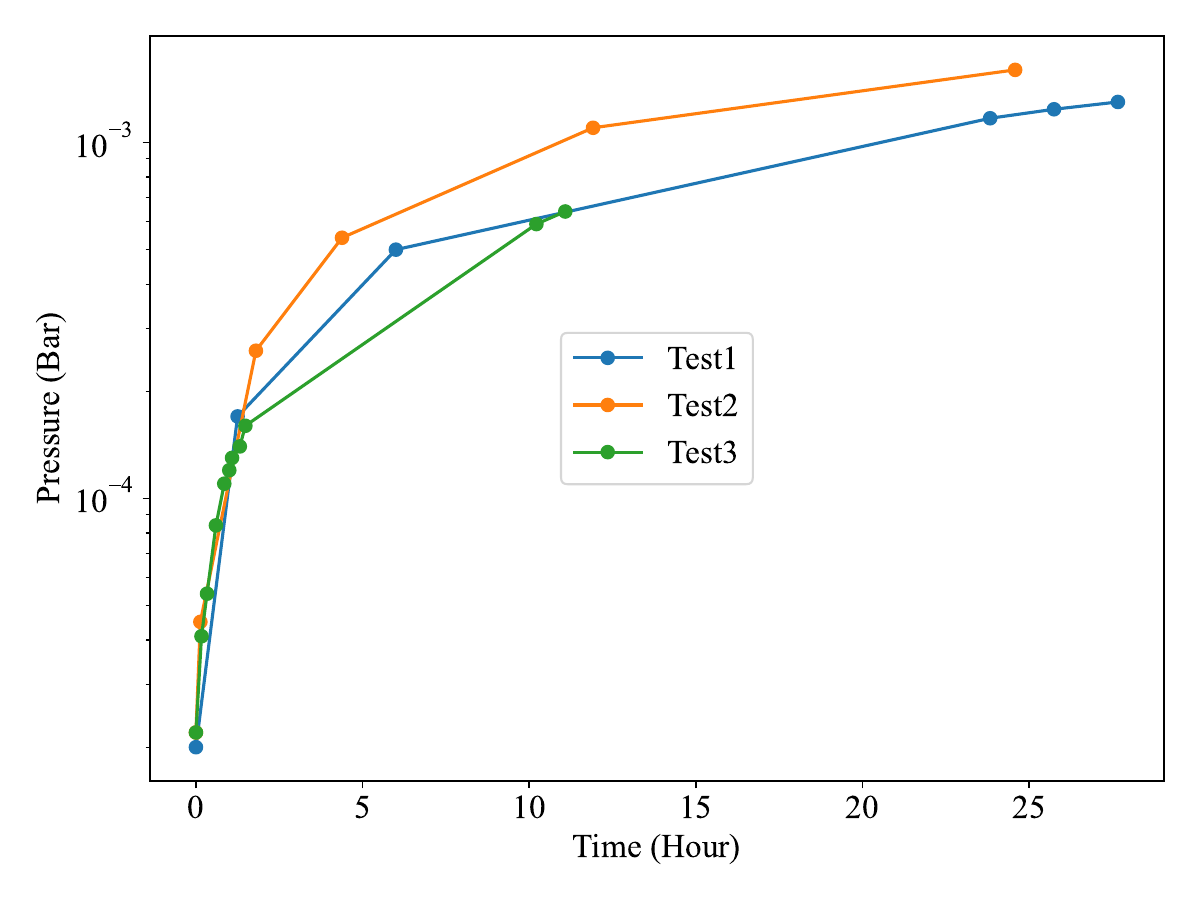} 
	%Here is how to import EPS art
	\caption{The pressure drops versus time elapses after the pump was shut off.}
	\label{figs3}
\end{figure}

{\bf Vibration noise.} Vibrations of the structure of the FMTO system introduce noise both in the dynamics of the FMTO and in laser spot displacement measurements. It is observed that the pressure in the vacuum chamber of the FMTO rises gradually over time. Typical values are shown in Fig.~\ref{figs3}. We found that the Q-factor of FMTO changed from approximately $10$ to $100$, while the Q-factor of a soft AFM cantilever (used as a reference) improved significantly from $10^2$ to $10^4$. Therefore, we conclude that although residual air in the vacuum chamber introduces some dissipation to the mechanical oscillator, it is not the primary factor affecting magnetic sensitivity in our current FMTO system. Consequently, the vacuum pump is switched off during experiments to reduce the vibrations.

{\bf Thermal noise.} 
For an FMTO with a quality factor of $40$ at a room temperature of $300\,\rm{K}$, the contribution from thermal noise to its PSD is approximately $4 \times 10^{-6}\,\rm{pxl^2\cdot Hz^{-1}}$ in the non-resonant condition, while it rises to approximately $7 \times 10^{-3}\,\rm{pxl^2\cdot Hz^{-1}}$ in the resonant condition. Notably, these results are $5$ orders of magnitude lower than the PSD of the FMTO, as depicted by the blue line in Fig. 2(a).

\section*{Supplementary Analysis 2: the capacity of FMTO}

For the FMTO system described in Eq.$(2)$, thermal noise arises from thermal
fluctuations in torque. According to the fluctuation-dissipation theorem,
the power spectral density of the thermal noise torque can be expressed as: 
\begin{equation}
	S_{\tau_{\rm{th}}}=4\left( k_{\rm{B}}T\right) I\gamma = 8\pi\left( k_{\rm{B}}T\right) I \,
	f_{\rm{r}}/Q,  \label{a03}
\end{equation}
where $T$ is the temperature, $I$ is the moment of inertia, $f_{\rm{r}}$ and $Q$
are resonant frequency and quality factor as given in the text, $k_{\rm{B}}$ is
the Boltzmann constant.

A magnetic signal can be detected when the torque generated by this signal
exceeds the thermal torque given by the above equation. The magnetic
sensitivity limited by thermal noise is therefore: 
\begin{equation}
	\eta_{\rm{B}}=S_{\tau_{\rm{th}}}^{1/2}/\mu ,  \label{a04}
\end{equation}
where $\mu$ is the magnetic moment of FMTO.

For a ferromagnetic sphere of radius $R$, mass density $\rho _{\rm{m}}$ and spin
density $\rho _{\rm{e}}$, it is noted that the magnetization $M=\rho _{\rm{e}}\mu _{\rm{B}}$%
, where $\mu _{\rm{B}}$ is the Bohr magneton. The moment of inertia is $%
I=2mR^{2}/5=\left( 8\pi /15\right) \rho _{\rm{m}}R^{5}$. When levitated in a
uniform magnetic field $B_{\rm{bias}}$, the magnetic moment of the sphere is $\mu
=4\pi R^{3}M/3=4\pi \rho _{\rm{e}}\mu _{\rm{B}}R^{3}/3$. The resonant frequency can be
derived with the equivalent elasticity coefficient $k=\mu B_{\rm{bias}}$, which
is 
\begin{equation}
	f_{\rm{r}}=\frac{1}{2\pi }\sqrt{\frac{k}{I}}=\frac{\sqrt{10}}{4\pi }\rho
	_{\rm{m}}^{-1/2}(\rho _{\rm{e}}\mu _{\rm{B}})^{1/2}B_{\rm{bias}}^{1/2}R^{-1}.  \label{a05}
\end{equation}%
This equation indicates that the resonant frequency increases with a
stronger magnetic field and decreases with a larger ferromagnetic sphere.
Two equivalent expressions for magnetic sensitivity, derived by substituting
the above expressions into the sensitivity formula, are: 
\begin{flalign}
	\eta _{\rm{B}} &=F_{\rm{0}}\left( k_{\rm{B}}T\right) ^{1/2}(\rho _{\rm{e}}\mu _{\rm{B}})^{-3/4}\rho
	_{\rm{m}}^{1/4}Q^{-1/2}B_{\rm{bias}}^{1/4}R^{-1},&  \label{a06} \\
	&=F_{1}\left( k_{\rm{B}}T\right) ^{1/2}(\rho _{\rm{e}}\mu _{\rm{B}})^{-5/4}\rho
	_{\rm{m}}^{3/4}Q^{-1/2}B_{\rm{bias}}^{-1/4}f_{\rm{r}},&  \label{a07}
\end{flalign}%
where $F_{\rm{0}}=(3/\pi )^{1/2}(2/5)^{1/4},F_{1}=\left( 2^{7}3^{2}\pi
^{2}/5^{3}\right) ^{1/4}$. \eqref{a06} illustrates the correlation between
the sensitivity and the radius of the FMTO, whereas \eqref{a07} shows the
relationship between the sensitivity and the resonant frequency of the FMTO.

\section*{Supplementary Analysis 3: probing exotic interactions}

A schematic diagram for probing exotic interactions using FMTO is shown in
Fig.3(b). The total pseudomagnetic field and the final integration result
are given in Eq.$(5)$ and Eq.$(6)$, respectively. Here, we outline the
detailed derivation process.

In the spherical coordinate system, the volume element is represented as $%
dV=\Omega r^{2}dr$, where $\Omega$ is the solid angle. This gives: 
\begin{flalign}
	\vecbr{B}^{\rm{pse}} & = f^{4+5}C_{\rm{0}}\int_{V}dV\left[ \left( \vecbr{v}_{\rm{n}}\times 
	\hat{r}\right) \left( \frac{1}{\lambda r}+\frac{1}{r^{2}}\right)
	e^{-r/\lambda}\right] ,  \notag \\
	& \approx  2\pi\varepsilon_{\rm{A}}\Omega C_{\rm{0}}f_{\rm{n}}\lambda^{2}f^{4+5}C_{\lambda
	}\left( l_{\rm{0}}\right) ,  &\label{a08}
\end{flalign}
where the $l_{\rm{0}}$ and $l_{\rm{m}}$ are the distances of the inner and outer
layers of the spherical shell-shaped nucleon source from the FMTO,
respectively. $C_{\rm{0}}=\hbar\rho_{\rm{n}}/(4\pi m_{\rm{e}}c\gamma)$ is a parameter
associated with materials and fundamental constants, with $\rho_{\rm{n}}$ the
nucleon spin density of the nucleon source, $m_{\rm{e}}$ the mass of electron, $\gamma$ the gyromagnetic ratio of electron, $\hbar$ and $c$ the reduced
Planck constant and the speed of light respectively.

The velocity amplitude $v_{\rm{n}}=2\pi f_{\rm{n}}A_{\rm{n}}$, with $f_{\rm{n}}$ the vibration
frequency and $A_{\rm{n}}=\varepsilon_{\rm{A}}l_{\rm{0}}$ the vibration amplitude of
nucleon source mass. The approximation in Equation \eqref{a08} is valid
because the contribution of the upper limit $l_{\rm{m}}$ is negligible compared $%
l_{\rm{0}}$ when $l_{\rm{m}}-l_{\rm{0}} > 7\lambda$. The function $C_{%
	\lambda}(l_{\rm{0}})=(l_{\rm{0}}^2/\lambda^{2}+2l_{\rm{0}}/\lambda)e^{-l_{\rm{0}}/\lambda}$ is a
dimensionless and reaches a maximum value of $1.174$ when $l_{\rm{0}}=\sqrt{2}%
\lambda$. This suggests the pseudomagnetic field is strongest near the
sensor when the distance from the source is $\sqrt{2}\lambda$, optimal for
probing interactions involving an unobserved boson with a Compton wavelength 
$\lambda$.

The results for probing exotic interactions that expressed by $\Delta f^{4+5}$
can be given:
\begin{flalign}
\Delta f^{4+5}&  =\frac{\Delta B}{\left\vert \partial B^{\rm{pse}}/\partial
		f^{4+5}\right\vert }=\frac{\eta_{\rm{B}}\left( f\right) T_{\rm{mea}}^{-1/2}} {%
		2\pi\varepsilon_{\rm{A}}\Omega C_{\rm{0}}C_{\lambda}\left( l_{\rm{0}}\right) f_{\rm{n}}
		\lambda^{2}},  \notag  \\
 &=\frac{\left( k_{\rm{B}}T\right) ^{1/2}\rho_{\rm{m}}^{3/4}F_{1}}{2\pi\varepsilon
		_{\rm{A}}\Omega C_{\rm{0}}C_{\lambda}\left( l_{\rm{0}}\right) \lambda^{2}\mu_{\rm{B}} ^{5/4}%
		\left[ Q^{1/2} \rho_{\rm{e}}^{5/4}B_{\rm{bias}}^{1/4}T_{\rm{mea}}^{1/2}\right] }.\label{eq11}&
\end{flalign}
In this equation, the absence of $R$ means that the sensor's radius has no
effect on the accuracy, whereas $\Delta f^{4+5} \propto \rho_{\rm{e}}^{-5/4}$
implies that the standard deviation is inversely proportional to the $5/4$
power of the spin density. Therefore, using materials with higher spin
densities for sensors can significantly improve the accuracy of probing
exotic interactions. Ferromagnetic materials, such as those with spin
densities up to $10^{28}\ \rm{{{{{{{m^{-3}}}}}}}}$, are particularly
suitable for such detection. Moreover, the sensor's radius have no impact on the accuracy of probing exotic interactions. Additionally, the FMTO system can
enhance its magnetic sensitivity and the precision of exotic interactions
measurements by a factor of $Q^{1/2}$ under resonance conditions($\Delta
f^{4+5}\propto Q^{-1/2}$). With a Q-factor up to $10^{7}$ for
superconducting levitated ferromagnetic spheres\cite{vinante2020}, FMTOs
have significant potential in magnetic sensing and probing exotic
interactions.

%\newpage 
\bibliography{sn-bibliography}